\documentstyle[epsfig]{article}

\begin{document}

\large{\center{\bf {Reply to ``Comments on Kullback-Leibler and
renormalized entropies: Applications to electroencephalograms of
epilepsy patients".}}}

\vspace{1cm}

{\center{\bf R. Quian Quiroga$^{1,2}$, J. Arnhold$^{2,3}$, K. Lehnertz$^3$ and P. Grassberger$^2$} \\

\vspace{0.5cm}

$^1${\sl Sloan-Swartz Center for Theoretical Neurobiology.}\\
{\sl California Institute of Technology, MC 216-76. 91125 Pasadena, CA. USA}\\
$^2${\sl John von Neumann Institute for Computing,}\\
{\sl Forschungszentrum J\"ulich GmbH, D - 52425 J\"ulich, Germany}\\
$^3${\sl Department of Epileptology, University of Bonn,}\\
{\sl Sigmund-Freud Str. 25, D - 53105 Bonn, Germany} \\
}

\vspace{1cm}

\begin{abstract}
Kopitzki et al (preceeding comment) claim that the relationship
between Renormalized and Kullback-Leibler entropies has already been
given in their previous papers. Moreover, they argue that the
first can give more useful information for e.g. localizing the
seizure-generating area in epilepsy patients.

In our reply we stress that if the relationship between both
entropies would have been known by them, they should have noticed
that the condition on the effective temperature is unnecessary.
Indeed, this condition led them to choose different reference
segments for different channels, even if this was
physiologically unplausible.
Therefore, we still argue that it is very unlikely that renormalized entropy
will give more information than the conventional Kullback-Leibler
entropy.
\end{abstract}

\newpage

We thank the authors of the preceding comment for pointing out a misprint in \cite{quian}
(in the line following Eq. (8) it should read $p\equiv \tilde{q}$ instead of $p\equiv q$),
and a numerical inconsistency in Figs. 2-4 of ref. \cite{quian}. The latter
resulted from an error in our code. The corrected data is shown in Fig. 1 below, where 
we added the renormalized entropy values calculated with a pre-seizure reference for
completeness.
Note that Eq. (10) of \cite{quian}, i.e. $| \Delta H | \leq K(p|q)$, is now
verified in all cases \cite{comment1}.
Despite this correction, the data are qualitatively similar
to the ones presented in \cite{quian}, and we still conclude that renormalized
entropy does not give more information than standard Kullback-Leibler (KL) entropy.

Apart from this, we do not agree with any of the other claims raised in the preceding
comment (and we still have some discrepancy in details with the numerical results shown
in \cite{kopitzki} whose origin is unclear to us).

The first main point of ref. \cite{quian} was to show that the ``renormalized entropy" (RE)
proposed in \cite{saparin} and applied to EEG data in \cite{kopitzki} was indeed a KL
entropy, but taking an unusual ``renormalized" reference.
We maintain, in contrast to claims made in the comment, that this relation (Eq. (9) in
\cite{quian}) was not mentioned in \cite{saparin,klimo-physica}, and not in \cite{kopitzki}
either. Indeed, due to it, the condition $T_{eff} \geq 1$ postulated in
\cite{saparin,klimo-physica,kopitzki} is not needed to obtain the inequality $\Delta H
\leq 0$. The fact that the latter was claimed in \cite{saparin,klimo-physica,kopitzki}
to hold only for $T_{eff} \geq 1$ indicates that the authors were not aware of the relation to KL 
(or ``relative") entropy. Apart from this, we also wanted to give a simple
treatment free of all allusions to statistical thermodynamics, the latter making the treatments
in \cite{saparin,kopitzki} hard to understand.

Our second point was that RE is very unlikely to be more useful than
the usual (un-renormalized) KL entropy for the analysis of EEGs from epileptic patients, as claimed
in \cite{kopitzki}. On the one hand this was based on the numerical similarity between
RE and standard KL entropies, which is enforced by several inequalities and which makes it unlikely a priori
that either is superior. On the other hand, we verified this explicitly by detailed
numerical calculations which indeed showed that both behaved very similar. 
It is clear from Fig. 1 that major differences are due to the choice of the 
reference window.
In contrast to what is suggested in the comment, we did not base this conclusion 
entirely on theoretical arguments.

Finally, we also stressed that the condition $T_{eff} \geq 1$ -- which is not
needed at all -- has led the authors of \cite{kopitzki} to choose reference points which
are physiologically very unfortunate.
Again, we remark that it is very unreliable to compare a relative entropy measure obtained
from EEG recordings at different electrodes by using different references (from 
the pre-seizure stage, during the seizure, or from the post-seizure stage) for each electrode in 
order to localize an epileptic focus.
Thus the failure of realizing that RE is a sort of
KL entropy -- or at least of drawing the obvious consequences from this observation -- has
hampered its application to EEGs.

\newpage
\begin{figure}
\begin{center}
\epsfig{file=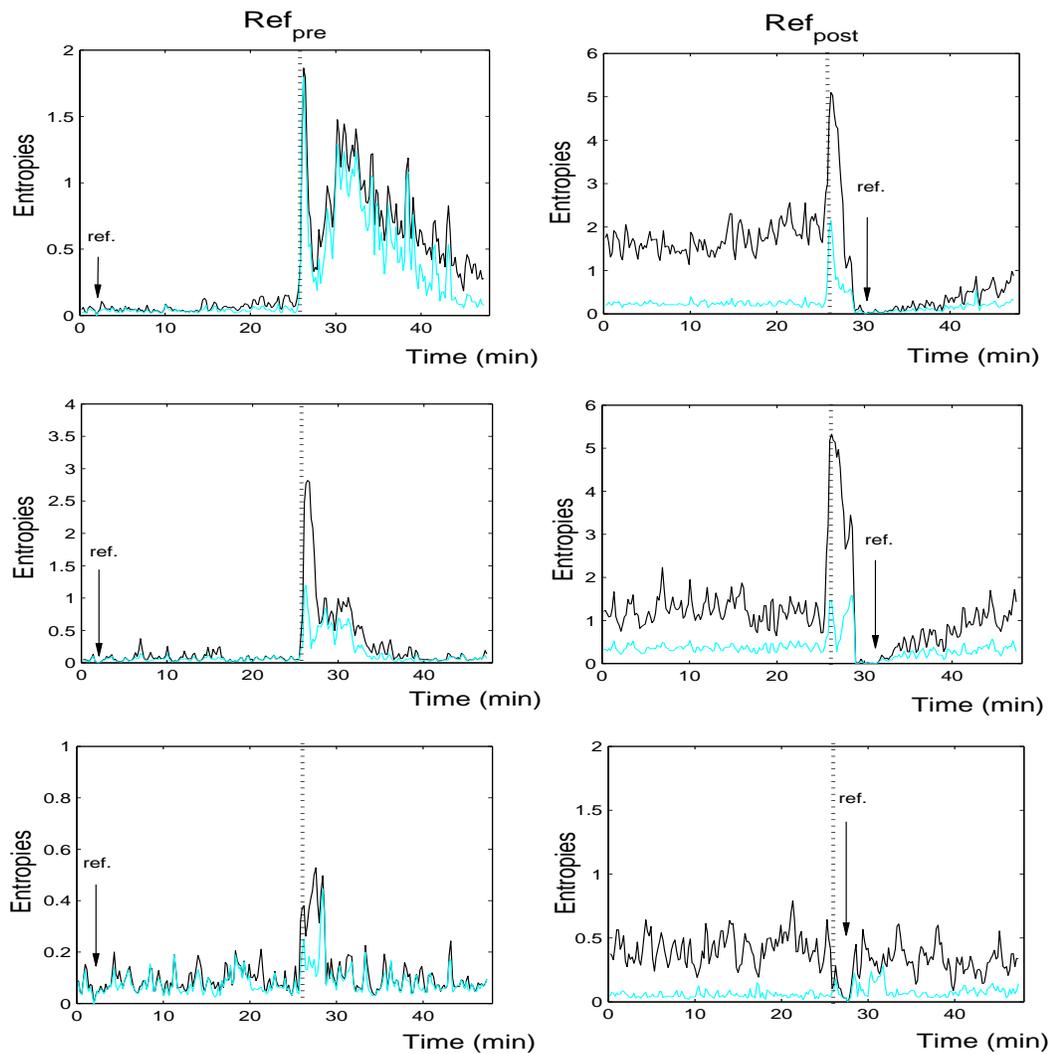,height=14cm,width=14cm,angle=0}
\end{center}
\caption{Kullback-Leibler (black line) and renormalized (gray line) entropies from EEGs recorded in the
seizure-generating area (upper row), adjacent to the seizure generating
areas (middle row), and in the non-affected brain hemisphere (lower row).
Data shown in left (right) columns were obtained from using a pre- (post-) seizure reference window (marked by an arrow). 
The dotted vertical lines mark the electrical onset of the seizure.}
\label{fig:1}
\end{figure}


\begin{thebibliography}{10}
\bibitem{comment1}
We do not understand how this relationship proved in ref.
\cite{quian} can be invalidated by ``basic mathematical
operations" as claimed by the authors in their reference [9].
\end{thebibliography}
\end{document}